Report on the ESO workshop

# Maintaining scientific discourse during a global pandemic: ESO's first e-conference #H02020

held virtually from 22 – 26 June 2020, daily between 12:50 – 15:10 UTC

Richard I. Anderson[1,2,3]
Sherry H. Suyu[4,5,6]
Antoine Mérand[1]

[1]ESO
[2]ORIGINS cluster, Universitätssternwarte München, Germany
[3]Institute of Physics, Laboratory of Astrophysics, Ecole Polytechnique Fédérale de Lausanne, Switzerland; richard.anderson@epfl.ch
[4]Max Planck Institute for Astrophysics, Garching, Germany
[5]Technical University of Munich, Garching, Germany
[6]Academia Sinica Institute of Astronomy and Astrophysics, Taipei, Taiwan

## Abstract

From 22 to 26 June 2020, we hosted ESO's first live e-conference, #H02020, from within ESO headquarters in Garching, Germany. Every day, between 200 and 320 researchers around the globe tuned in to discuss the nature and implications of the discord between precise determinations of the Universe's expansion rate, $H_0$. Originally planned as an in-person meeting, we moved to the virtual domain to maintain strong scientific discourse despite the SARS-CoV-2 (COVID-19) pandemic. Here, we describe our conference setup, participants feedback gathered before and after the meeting, and lessons learned from this unexpected exercise. As e-conferencing will become increasingly common in the future, we provide our perspective on how e-conferences can make scientific exchange more effective and inclusive, in addition to climate friendly.

## Before 18 March 2020: in-person conference at ESO HQ Garching

Our preparations for an in-person conference involving approximately 100 participants began in Summer 2019, shortly after the KITP Conference on Tensions between the Early and the Late Universe (Verde et al. 2020). Together with the Scientific Organizing Committee (SOC), composed of Chuck Bennett, Annalisa Calamida, Matthew Colless, Frédéric Courbin, Claudia de Rham, Wolfgang Gieren, Chow-Choong Ngeow, Hiranya Peiris, Mickael Rigault, Dan Scolnic, and Licia Verde, we selected invited speakers that would cover a wide range of topics from basic astrometry to theoretical cosmology. The conference was planned from Monday afternoon to Friday at noon, featuring 20 invited talks, 2 introductory lectures, 34 contributed talks, 8 short 10-minute discussion sessions, four 10-minute poster flash talk sessions in addition to physical posters, and a 1½ hour discussion panel at the end. A public lecture by Adam Riess at the Technical University Munich's downtown campus was also planned.

Registration for the workshop opened in late January 2020 with abstracts due 1 April. However, as cases of COVID-19 started skyrocketing in February and March, it eventually became clear that an in-person meeting in June was unrealistic. Faced with the question of whether to postpone or cancel the meeting, on 26 March we decided for a third option: to convert the in-person meeting to a live global e-conference within the course of 12 weeks.

We had no blueprint to follow and no pre-arranged technical solutions for implementing ESO's first global live e-conference, leading to a somewhat tricky situation that was rendered even less straightforward by nearly all ESO staff working remotely for the first time. Nonetheless, we strongly felt that maintaining strong international scientific discourse was worth the challenge – especially since many other meetings were being canceled or postponed indefinitely – and decided to go ahead with this meeting despite the short planning timescale.

## Participant surveys before & after the meeting

Recognizing the experimental nature of our e-meeting, we conducted participant surveys before and after the meeting to assess how well our meeting met participant needs and expectations, and to measure whether the meeting fulfilled our own goals. Before the conference, we collected 89 responses from unique individuals, and 79 afterwards. Since anonymous submissions were allowed, only 46 before & after responses could be uniquely matched. The following account is based on this survey information.

## Goals of the online meeting

In adapting the conference for the virtual domain, we pursued the following main goals:

1. To advance the specific research field by enabling continued international scientific exchange
2. To create a schedule compatible with most regions of the world
3. To foster informal discussions that go beyond the scope of the invited presentations
4. To strike a balance between giving the wider community a strong voice and covering a broad range of topics via invited presentations
5. To explore and leverage benefits inherent to online meetings, such as reduced access barriers to foster diversity, equity, and inclusion
6. To provide inspiration and guidance for those considering to host an e-conference

The meeting setup was chosen such as to maximize the above goals, to the possible detriment of other worthwhile goals, such as networking sessions or online poster sessions focusing on young researchers, which we unfortunately could not include due to the compactified planning schedule.

## Meeting setup

Figure 2 illustrates the e-conference setup adopted, which was also explained in a YouTube video[1].

Targeting Goal 2, we selected the time slot of 12:50 – 15:10 UTC, every day from Monday, 22 through Friday, 26 June 2020, which is at least somewhat close to normal waking hours in most areas of the world. However, especially the start (5:50am in Vancouver) and end times (1:10am in Canberra) were somewhat uncomfortable for regions bordering the Pacific Ocean. No breaks were included in the short 2h20 minute window. Feedback indicated overall approval of this time slot. Some participants found breaks missing while others thought it appropriate to skip breaks since most were following the meeting from home.

Each day's session consisted of four invited talks (20 minutes talk + 5 minutes Q&A) and a 30-minute live discussion panel[2]. The latter served dual purposes of providing a voice to the community (in lieu of contributed talks) and enabling critical discussion, thus targeting Goals 3 and 4. The panels featured prepared statements, informal group discussion, and addressed questions left unanswered after the invited talks. Panels were composed of the day's invited speakers and 3 to 5 participants who had requested to be panel members during registration.

The conference call was held on Zoom[3] and live streamed via YouTube[4] (incurring a ~20s delay), where all videos remain publicly accessible and have been viewed more than 5,000 times. Only invited speakers, discussion panelists, and moderators were invited to unmute their microphones in the Zoom call to ensure an orderly and uncomplicated meeting. To add a human touch, each day's session began and ended with a brief greeting during which all Zoom participants were asked to turn on their video and greet each other. To provide a lively response to each talk, moderators thanked speakers by applauding. "Canned applause" was suggested by Tom Shanks as a suitable replacement.

Planning and adhering to a tight schedule was crucial to ensure a smooth meeting involving live participants across 18 time zones. We therefore did extensive onboarding work with invited speakers and provided a short (4m46s) YouTube video explaining the meeting setup to all participants. Onboarding was done in one-on-one Zoom calls during which we explained the meeting setup, launched mock presentations, and answered any technical questions. We believe that this onboarding work was the reason why few technical issues were experienced during the conference. Although we briefly considered pre-recording talks to minimize technical issues, we decided in favor of live presentations to preserve the more direct feeling of a real conference and to avoid speaker stress of recording “the perfect talk”.

---

[1] https://www.youtube.com/watch?v=j2HkdQpc7fc&t=1s
[2] https://www.eso.org/sci/meetings/2020/H0.html
[3] https://www.zoom.us
[4] https://www.youtube.com

Live participation on Zoom and YouTube (up to 330 live participants on Day 1) exceeded our in-person expectations three-fold, while asynchronous streaming from YouTube has reached a 10 times greater audience than a conventional meeting would have.

All questions were submitted via the online platform Slido[5], which allowed participants to upvote relevant questions and supported anonymous question submission. Figure 1 shows a word cloud created from the questions submitted. Questions deemed relevant by a majority were then relayed to speakers by the session moderator. Advantages of Slido included very concise formulations due to the 160 character limit, democratization of Q&A thanks to voting, the ability to ask questions anonymously, and a written account of all questions, which allowed speakers to reply in writing to any unanswered questions. A simultaneous benefit and drawback of using Slido for Q&A was that follow-up questions were not possible; the advantage being that precious Q&A time could not be exhausted by individual questions and the disadvantage being that some questions were not satisfactorily answered. Post-conference feedback shows that a large majority of participants were extremely pleased with this form of Q&A.

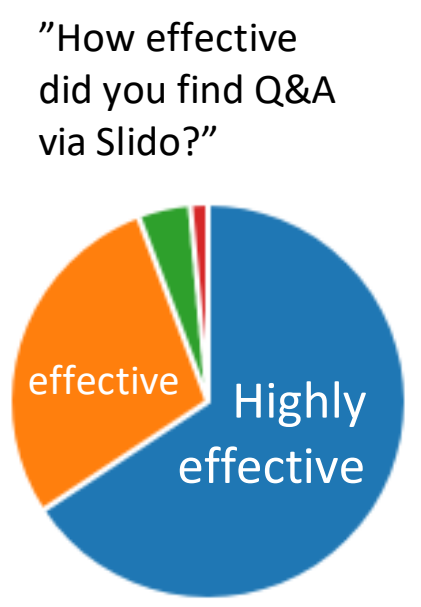

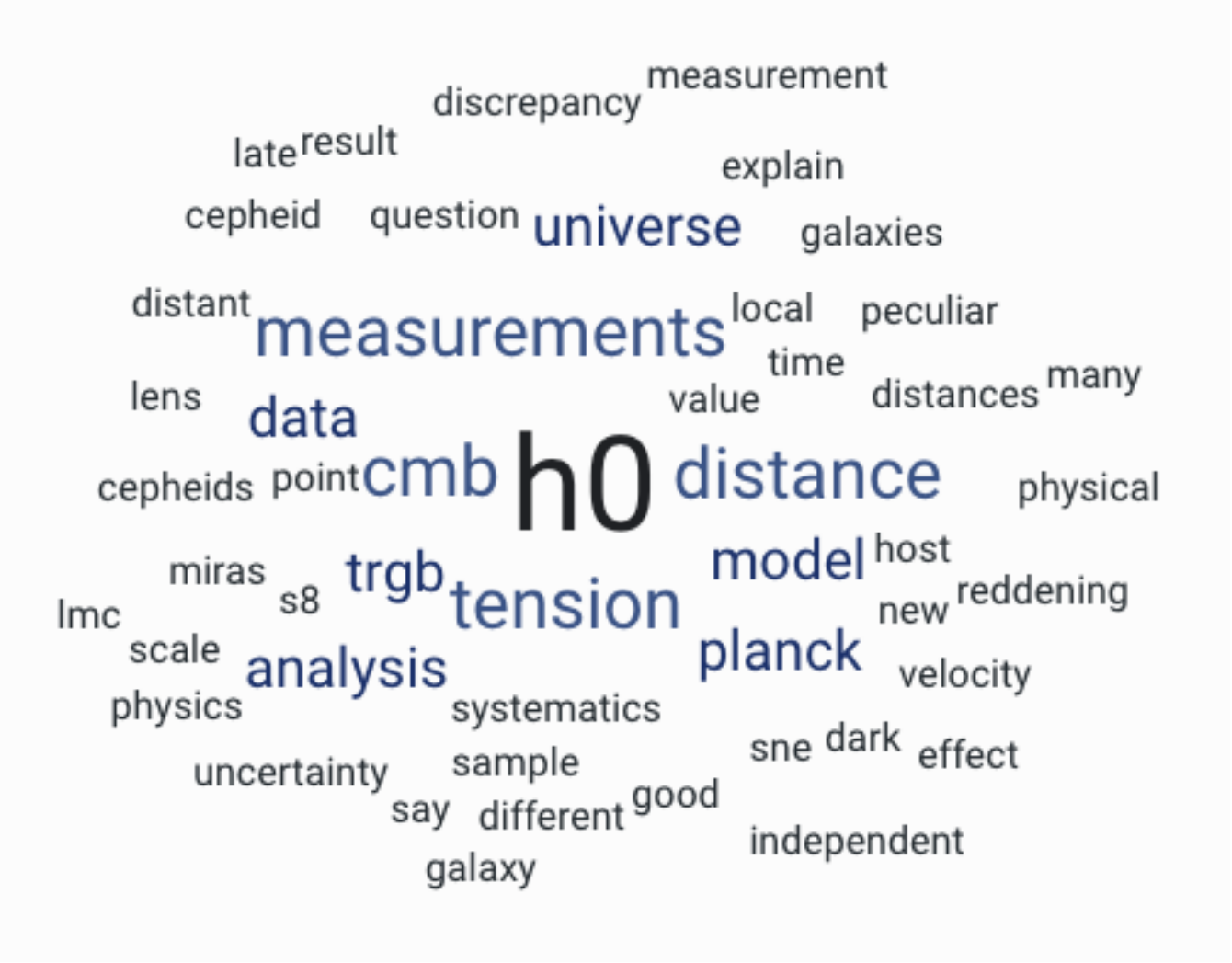

*Figure 1: Word cloud created from questions on Slido. 301 questions were asked, 48% of them anonymously, and 1168 likes were given for voting.*

We created a Slack workspace[6] to support asynchronous exchanges for this event. In principle, Slack allows the exchange of information in chat channels consisting of various user groups as well as to get in touch directly with each other via one-on-one chat or video calls. Over the course of the conference week, participants exchanged more than 1500 messages, including also organizational messages, scientific discussions, answers to Q&A, diagrams and other documents, such as presenter slides. However, becoming familiar with several new communication tools at once (Zoom, Slido, Slack) represented an initial hurdle for some participants. Yet, once accustomed, participants reacted very positively to Slack as a coherent communication platform for the meeting.

[5] https://sli.do

[6] https://h02020.slack.com

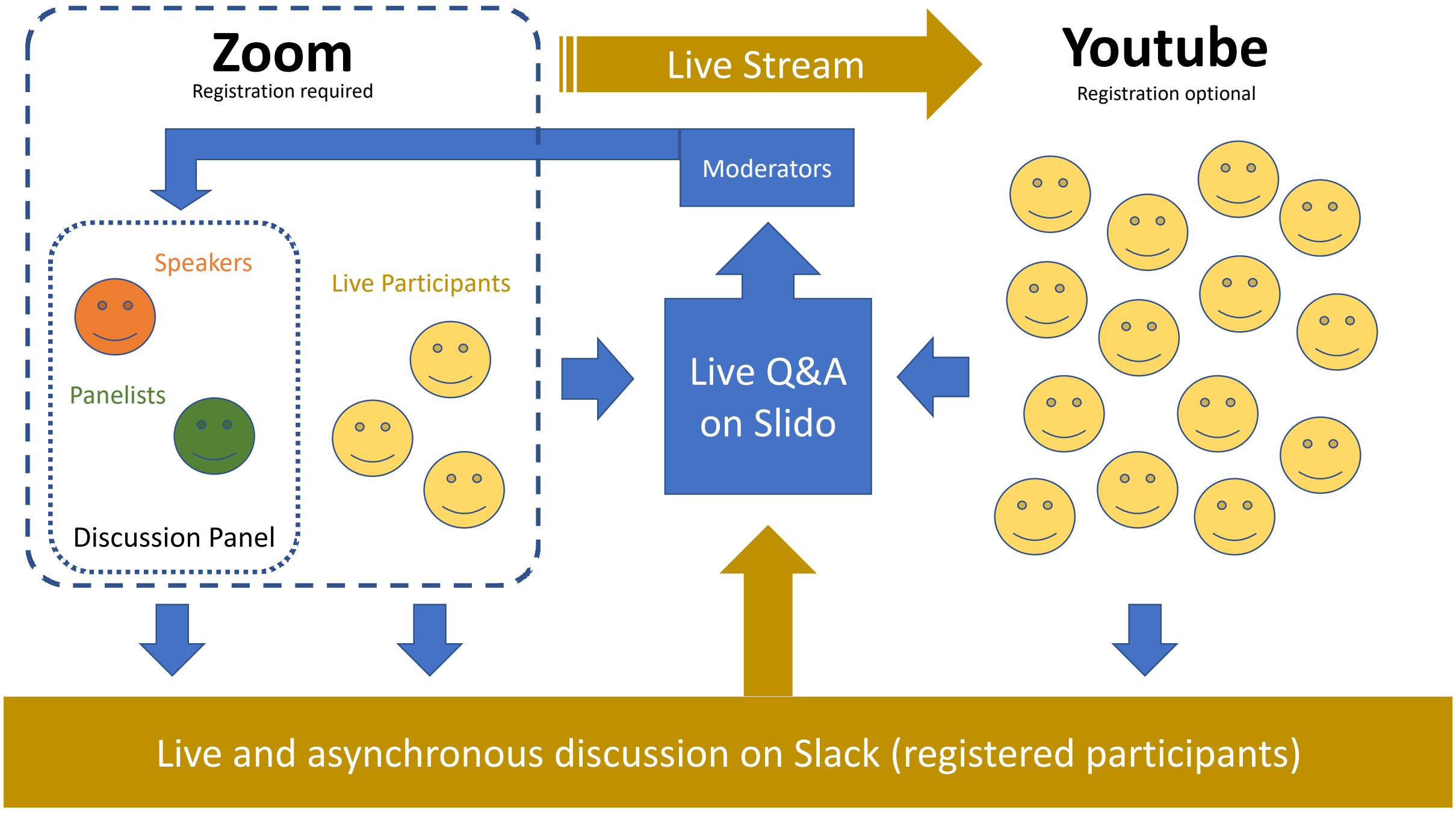

Figure 2: Meeting setup

## Participation & representation

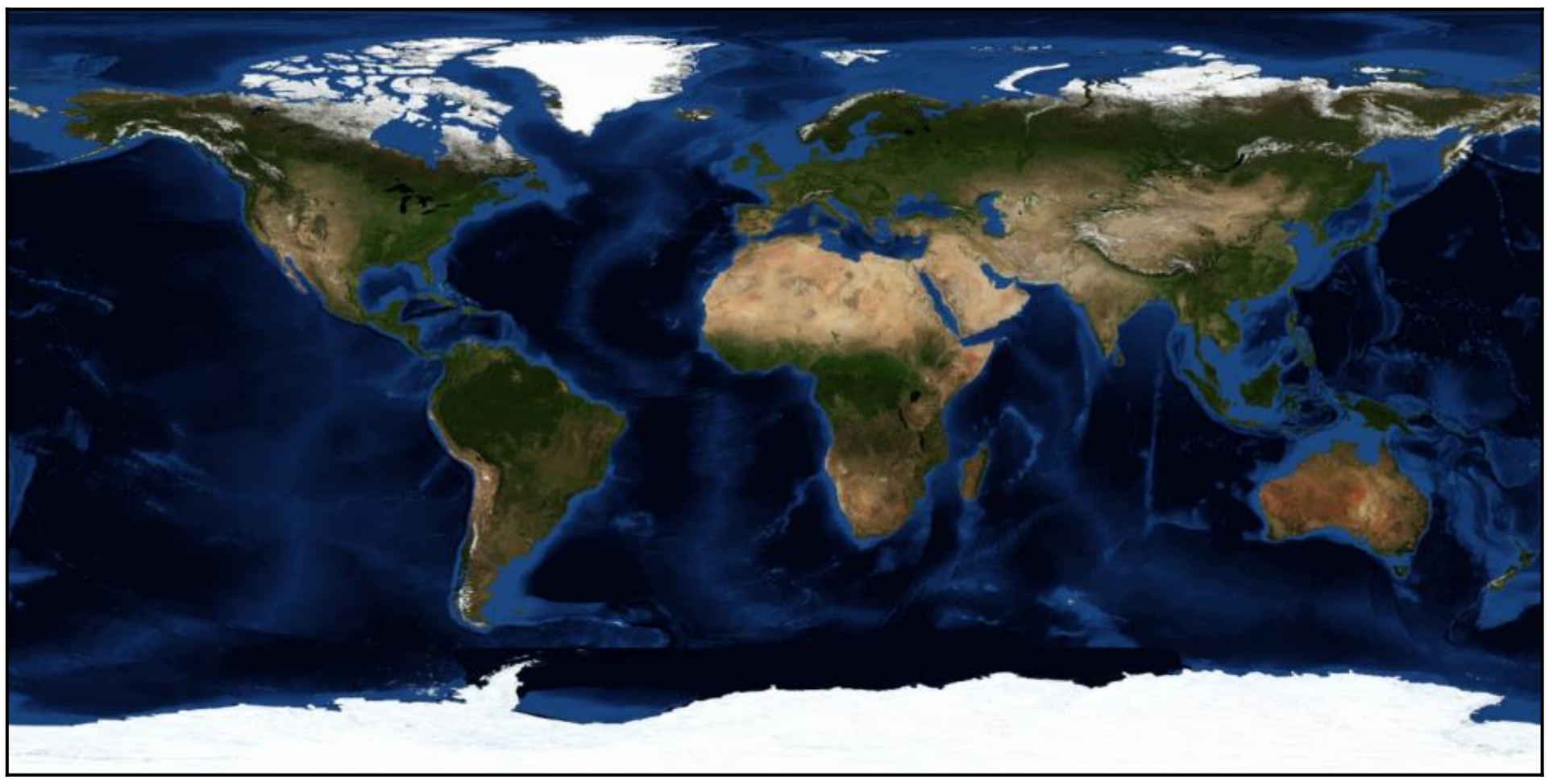

Figure 3: Registration locations by IP address

*Table 1: Career and gender demographics. Gender was assigned according to first names provided. One postdoc's name could not be assigned a gender.*

| Registered Participants | N total | % total | N Female | % Female | N Male | % Male |
|---|---|---|---|---|---|---|
| All | 295 | 100 | 79 | 27 | 216 | 73 |
| Student | 87 | 29 | 33 | 38 | 54 | 62 |
| Postdoc | 62 | 21 | 16 | 26 | 45 | 73 |
| Staff | 64 | 22 | 16 | 25 | 48 | 75 |
| Professor | 70 | 24 | 13 | 19 | 57 | 81 |
| Emeritus/retired | 6 | 2 | 0 | 0 | 6 | 100 |
| Teacher | 1 | | 0 | 0 | 1 | 100 |
| Journalist | 1 | | 0 | 0 | 1 | 100 |
| Freelance researcher | 1 | | 0 | 0 | 1 | 100 |
| Other | 3 | 1 | 0 | 0 | 3 | 100 |
| | | | | | | |
| SOC | 14 | | 5 | 36 | 9 | 64 |
| Invited speaker | 20 (6 early career) | | 7 | 35 | 13 | 65 |
| Panel members (by request) | 22 | | 5 | 23 | 17 | 77 |

We sought fair community representation among SOC members, who helped select a diverse set of invited speakers across the spectrum of career stages, locations, gender, and subject matter. Registration was open until the 300 person limit of the Zoom license was exhausted. The only information collected concerning the identity of participants was name, position, and affiliation. Gender statistics shown in the table above are derived from manually (and possibly incorrectly) assigned (binary) gender based on first names. The world map in Figure *3* was created using the IP addresses of participants during registration.

Representation of women among invited speakers and the SOC was significantly higher than among all registered participants. Gender representation among panel members mirrored the registration demographics. Two participants who had requested panel slots were not assigned because they either misunderstood the panel setup or because they had not yet completed graduate studies. Six of 20 invited speakers were early-career researchers.

We note the sharp decrease in female representation between the categories students and postdocs. It is a well-known fact that retaining qualified women is a major issue in STEM fields, and the observed trend suggests that women are less likely than men to continue beyond their doctorate degrees towards a career in astronomy. However, we do not have sufficient information to assess this drop-off in detail, or to cross-reference it with other potentially relevant factors, such as postdoc-seniority or child care responsibilities, which have disproportionately impacted women during the COVID-19 pandemic.

Representation by people of color among speakers and panelists was low, especially counting non-Asian people. This contrasted the timing of the meeting, which coincided with #BlackinSTEM week. There are many and complex reasons for underrepresentation of people of color in STEM, and they may differ significantly across different regions of the world. However, especially online conferences should strive to do better since barriers for participations (funding, travel, etc.) are lower than for in-person conferences. We note that we could not identify any participants from the African continent based on either IP address, e-mail address, host institute name, or YouTube analytics data, although we did have participants from several underpriviledged countries. We had listed our conference on the CADC website[7] and advertised it through ESO's mailing list, newsletter, and website, several astronomy-related mailing lists, and began advertising it on Twitter. We are concerned that our announcement practices could have been exclusionary to people from certain regions, and we have identified that internet censorship would have prevented potential participants from China or Iran from seeing our Twitter posts or YouTube live streams. Professional organizations such as the IAU could play an important role in providing guidance on best practices for increasing diversity in virtual conferences.

Up to 120 participants (registered or not) joined via the YouTube livestream every day. YouTube channel analytics provide information concerning audience age, location, and gender as shown in Table 2 below. The analytics available are unfortunately not very detailed and likely based on inhomogeneous samples including registered and unregistered YouTube users, depending on category. For example, the total number of views by location adds up to merely 46%; location information is missing for the majority of viewers. Nonetheless, we see that YouTube was favored by a younger audience, possibly due to different approaches to using the internet and/or different career stages.

[7] https://www.cadc-ccda.hia-iha.nrc-cnrc.gc.ca/en/meetings/

*Table 2. YouTube channel analytics*

| YouTube participation | % views | % watch time |
|---|---|---|
| Female | 22.4 | 23.5 |
| Male | 77.6 | 76.4 |
| English Subtitles | 7.6 | 13.0 |
| | | |
| Age group 18-24 | 4 | 4 |
| Age group 25-34 | 72 | 70 |
| Age group 35-44 | 22 | 24 |
| Age group 45-54 | 3 | 2 |
| Total North America | 15.3 | 14.1 |
| US | 13.8 | 12.7 |
| Total South America | 1.1 | 1.1 |
| Chile | 1.1 | 1.1 |
| Europe | 28.9 | 34.4 |
| Germany | 10.4 | 12.6 |
| France | 7.4 | 5.9 |
| United Kingdom | 5.7 | 8.6 |
| Asia | 0.2 | 0.2 |
| Australia | 0.8 | 0.8 |

### Goal 5: reducing access barriers to increase diversity, equity, and inclusion

We deliberately collected no registration fees to allow any interested parties to participate. For comparison, the in-person registration fee would have been 180 EUR (80 EUR students). After converting our conference to an online format, we set up a new registration form. The registration served several purposes, including limiting access to the Zoom call and Slack workspace, collecting consent to being recorded/live streamed, and collecting the pledge of participants to abide by the ESO code of conduct. However, the full conference was publicly accessible even without registration via the YouTube live stream and questions on Slido. All software tools used (Zoom, Slido, Slack) were free of charge to participants and compatible with a maximum of operating systems with no installations required apart from a WebRTC capable browser (e.g. Firefox, Chrome, Opera).

Most feedback mentioned that removing the need to travel was a key advantage of the virtual format, irrespective of the Corona pandemic. Specific reasons given included child care, teaching and other work-related duties, care of pets, visa-related issues, time saved by not travelling, avoiding exhaustion (e.g. jet lag), health concerns, dietary considerations, and more. An Australian participant emphasized that these benefits far outweighed the inconvenience of late night sessions.

Anonymous questions were perceived by many as an important advantage over classical Q&A. With no need to fear embarrassment, many basic questions were submitted and voted for, in particular during more theory-heavy talks. Irrelevant questions – a concern for those opposed to anonymous questions – were not an issue because they were very few and did not get upvoted. The 160 character limit of Slido was considered a challenge by some participants, although lengthier questions could also be relayed to the Slack workspace.

A notable benefit of YouTube was the ability to enable closed captioning (English subtitles). The average view duration was 38:22min with subtitles enabled, compared to 20:59min without subtitles; the 7.6% of views with subtitles enabled accounted for 13% of the time watched on YouTube. This underlines the need to assist participants in engaging with the materials presented, especially in the case of persons with impaired hearing or non-native English speakers. George Jacoby (NOIRlab) further pointed out that participation conditions were more equal here than in usual meetings because everyone sat in the front row with unobstructed view and adjustable speaker volume.

A frequently mentioned drawback of the virtual format is that it is less conducive to informal discussions than in-person meetings. However, survey feedback also revealed interesting potential benefits to those discussions that did take place on Slack. First, discussions on Slack were transparent to all participants instead of only a small group of people (e.g. coffee break clusters), helping non-specialists gain deeper insights “behind the scenes”. Second, discussions could later be synthesized from the recorded chat text. Third, one participant mentioned they felt more at ease entering into an online discussion with strangers than they would have in person.

## Overcoming shortcomings

The ability for young scientists to present themselves and their work is crucial to foster their career development and to provide a forum for “hot-off-the-press” results that may revolutionize a field in the future. Unfortunately, we were not able to arrange for contributed talks, e.g. by students and early career scientists, in this instance. However, feedback offered two very attractive options for including early-career contributions despite a tight live schedule: a) pre-recorded contributed talks available for asynchronous viewing ahead of invited talks and discussion sessions; b) online poster sessions on Slack. To prevent exclusion by internet censorship, all essential content should be rendered accessible on platforms freely accessible worldwide without censorship.

While e-conferences are not subject to travel-related access barriers, several other barriers may apply to the same groups who would have difficulty attending in-person meetings. These include: internet censorship, lack of broadband internet access, the need for access to suitable personal devices (e.g. 1 laptop per person), among other aspects that should be considered even before planning virtual meetings. One way of addressing such issues could be to create regional viewing hubs once health measures allow it, cf. Reshef et al. (2020).

Perhaps the most common negative feedback from participants was that they had wished for more opportunities to discuss. The key to improving this aspect of e-conferencing seems to be to motivate participants to commit to offline discussions. Obstacles to this end include a lack of engagement (intentional or circumstantial) and the complexity of orienting oneself in a Slack workspace. During #H02020, we witnessed participants become increasingly engaged as they learned how to use this tool. In particular the ability of direct video calls among Slack workspace members seems to have been underused. Instructions for how to use Slack should thus be part of the onboarding information, and using the same tools in all conference-related communications would lower the need for participants to familiarize with new tools each time. Additionally, asking participants to specify keywords upon registration could help to assign discussion groups, connect participants according to interests, and foster networking.

## Final thoughts and recommendations

The fact that e-conferencing is much more climate-friendly than classical conferences is widely known in the community and the carbon savings of e-conferences have been described in detail, e.g. by Jahnke et al. (2020). Nonetheless, it took a worldwide health emergency to accelerate the adoption of e-conferencing. Now that the initial hurdle has been taken, e-conferences are becoming commonplace and will likely become an integral part of scientific discourse, not least because they are cheaper and more convenient for participants. Targeting Goal 6, we now close with some final thoughts and recommendations.

First and foremost, feedback clearly shows that participant satisfaction with this conference (both on the scientific and technical level) was high, and it is worth noting that no one stated that the conference had been worse than expected. Instead, many were positively surprised, and a majority stated that this meeting increased the likelihood of them organizing an e-conference themselves. Especially the Q&A sessions on Slido were a highlight for most, with 58% (46/79) saying Q&A was very effective, 25% somewhat effective, and only 1/79 indicating that it was somewhat ineffective. Several people even stated they hoped for Slido-style Q&A to be incorporated in future face-to-face meetings.

”Did this event make it more likely for you to host an e-conference?”

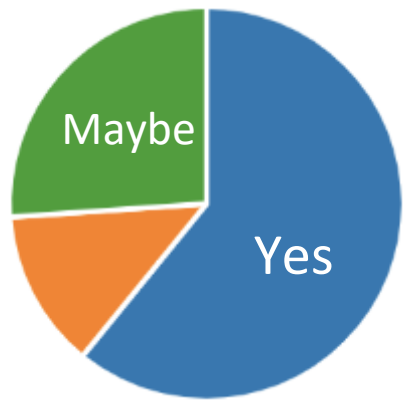

Of course, valid criticisms exist about e-conferences and their limitations, notably concerning the lack of direct human interactions. Participants especially noted the following issues, in arbitrary order: a) lack of body language, b) difficulty meeting new people, c) missing off-the-record discussions, d) the missing cultural elements of international travel that are crucial to foster understanding across cultures, languages, etc., e) missing networking opportunities, f) missing opportunities for senior and junior researchers to meet. Some of these drawbacks that may seem unsurmountable now may be addressed by smart e-conference design or may evolve as social norms evolve, for example regarding (online) networking etiquette. However, the human aspect of conferences is vital to scientific exchange and must not be neglected.

E-conferences may also harbor long-term negative side effects. For example, privacy concerns remain underdiscussed despite recorded live streams drastically changing the dynamic and persistence of participant contributions. Additionally, e-meetings tend to shift costs and efforts of attending conferences from the professional to the private domain, e.g. regarding food & drink, physical (office) space, computing resources, etc.

Yet, classical in-person conferences also have significant shortcomings that are easily forgotten or overlooked, perhaps because the community is used to them. For example, classical conferences are biased both implicitly (e.g. stereotypes, personalities) and explicitly (e.g. ability to travel), resulting in unintended exclusionary practices or situations. As a result, in-person meetings prioritize specific types of human interactions over other interactions that could lead to other benefits, such as transparent discussions, and foster equity, diversity, inclusion.

Social media are the native communication platform for e-conferences. Organizers should include social media in their announcement strategies and seek advice from press or outreach offices to make adequate use of social media, bearing in mind the possibility of censorship.

Another issue to solve is that of conference software complexity. At the time, our review of software available revealed no single solution that would have incorporated all the functionality desired for our meeting, so that we adopted a hybrid approach involving Zoom/YouTube, Slack, and Slido. Although corporate solutions do exist, they can be costly and were thus not applicable for an event with zero registration fees. A detailed comparison between different software solutions is out of scope of this article, but should certainly be conducted. The documentation and research gathered by the *EuroPython conference series*[8] offers substantial documentation concerning the technical aspects of online conferencing.

It is not clear that this conference could have taken place perhaps as little as 5 years ago. Projecting this rapid progress into the future, new tools and solutions will become available that may lead to even more options to choose from, and thus more complexity for conference attendees. To facilitate virtual-domain scientific cooperation in the future, it may be useful to define or develop a platform or set of tools (perhaps an open source version of Slack hosted by an institute) that the community can adopt over a time span and that renders the learning curve worthwhile. While the creative and competitive development of new tools certainly leads to innovation, standardization would make it worthwhile for scientists to invest time in becoming familiar with such channels.

We believe that e-conferencing can replace a majority of in-person conferences in the future, and we see many good reasons for this development. These notably include environmental benefits, time savings, opportunities for increasing diversity, equity, and inclusion, as well as the ability to create new conference formats that allow to improve global scientific discourse. Yet, the human element of scientific exchange must also be satisfied for science to progress. Hence, in-person meetings will remain a key aspect of scientific exchange. Combining the two options would thus benefit from the best of both worlds, and we believe this could be achieved by increasing coordination among conference organizing committees. For example, one could imagine a few very large international meetings per year that primarily focus on networking and personal interactions[9]. These in-person networking events could be complemented by frequent e-conferences focused on presenting scientific results, with the possibility for networking events centered around regional screenings. At the same time, free online seminar series, such as the Golden Webinar series[10] or the ESO Cosmic Duologues[11], are offering new opportunities for scientific exchange.

The scientific community has entered a new era of possibilities for scientific exchange. We argue that there could be immense benefits on the horizon, and reducing carbon footprint is certainly one of them. However, the drawbacks, challenges, and potential dangers of e-conferencing should not simply be ignored. A larger conversation should therefore consider how e-conferencing will become a safe, inclusive, and carbon friendly addition to the landscape of international scientific discourse.

---

[8] *https://www.europython-society.org/post/617463429296472064/sharing-our-research-and-licenses-for-going-online*
[9] Once traveling is no longer a health issue
[10] https://www.uc.cl/en/news/golden-webinars-in-astrophysics
[11] https://www.eso.org/sci/meetings/garching/Cosmic-Duologues.html

## Acknowledgements

We thank all the participants of #H02020, and especially the speakers, panelists, SOC members, and the many participants who provided feedback!

We thank Stella Chasiotis-Klingner and Nelma Alas Da Cunha Dias Da Silva for their administrative assistance, ESO safety and logistics personnel for helping us set up the H02020 control center in a safe environment during a health pandemic, and Marius Chelu for setting up the room infrastructure.

Many colleagues showed enthusiasm for an e-conference and contributed ideas that made this meeting a success. These include (in no particular order): Jason Spyromilio, Michael Hilker, Giacomo Beccari, Remco van der Burg, Marianne Heida, Dominika Wylezalek, Anna Miotello, Carlo Manara, Sara Mancino, ESO Director for Science Rob Ivison, ESO Director General Xavier Barcons, and many others. We also thank Behnam Javanmardi for covering the conference on reddit[12].

We thank the Max-Planck-Institute for Astrophysics (MPA) for providing access to their Zoom license, and Andreas Weiss at MPA for IT support. We thank Slack and Slido for offering free trial versions of their platforms, which made this conference setup possible, and YouTube for offering the free live streaming capability as well as the ability to revisit videos later.

We thank the Technical University of Munich for providing the venue for the public talk that was originally planned for the in-person conference, and Karin Lichtnecker and Gudrun Obst for the arrangements.

RIA acknowledges support through an ESO fellowship, the ORIGINS excellence cluster visitor program, and from the Swiss National Science Foundation through an Eccellenza Professorial Fellowship (award No. 194638). SHS thanks the Max Planck Society for support through the Max Planck Research Group, and the European Research Council (ERC) for support under the European Union's Horizon 2020 research and innovation program (LENSNOVA: grant agreement No 771776).

# References

Jahnke et al. 2020, Nature Astronomy, Volume 4, p. 812-815 (2020); https://doi.org/10.1038/s41550-020-1202-4

Reshef, O., Aharonovich, I., Armani, A.M. *et al.* How to organize an online conference. *Nat Rev Mater* **5,** 253–256 (2020); https://doi.org/10.1038/s41578-020-0194-0

Verde, Treu & Riess, Nature Astronomy, Volume 3, p. 891-895 (2019); https://doi.org/10.1038/s41550-019-0902-0

12 *https://www.reddit.com/r/cosmology/comments/hdwjgp/eso_conference_h0_2020_assessing_uncertainties_in/*